\documentclass[%
 aip,
 jmp,%
 amsmath,amssymb,
 reprint,%
]{revtex4-1}

\usepackage[mathlines]{lineno}
\usepackage{amsmath,amssymb,amsthm}
\usepackage{graphicx}

\newcommand{\al}{\alpha}
\newcommand{\be}{\beta}
\newcommand{\de}{\delta}

\newcommand{\ga}{\gamma}

\newcommand{\la}{\lambda}

\newcommand{\si}{\sigma}

\newcommand{\Ga}{\Gamma}
\newcommand{\La}{\Lambda}

\newcommand{\Om}{\Omega}
\newcommand\hg{\widehat{g}}
\newcommand\Bg{\bar{g}}
\newcommand\Bga{\bar{\ga}}

\def\BB{\mathbb{B}}

\def\RR{\mathbb{R}}

\newcommand\es{{\mathbb{S}^{n-1}}}
\newcommand\ball{\mathbb{B}^{n}}
\newcommand\oball{\overline{\mathbb{B}^{n}}}

\newcommand{\cE}{{\mathcal E}}
\newcommand{\tcE}{\widetilde{\mathcal E}}

\newcommand{\cF}{{\mathcal F}}
\newcommand{\tcF}{{\widetilde{\mathcal F}}}
\newcommand{\cG}{{\mathcal G}}

\newcommand{\cO}{{\mathcal O}}

\newcommand{\pd}{\partial}
\newcommand\minus\backslash

\renewcommand\leq\leqslant
\renewcommand\geq\geqslant
\newcommand\less\lesssim
\newcommand\gr\grtsim

\newlength{\intwidth}

\newcommand\AdS{{\mathrm{AdS}_{n+1}}}
\newcommand\gads{g_{\mathrm{AdS}}}
\newcommand\gref{\ga_0} 

\newcommand\bgads{\bar g_{\mathrm{AdS}}}
\newcommand\gcyl{g_{\RR\times\es}}

\newcommand{\triple}[1]{{\left\vert\kern-0.25ex\left\vert\kern-0.25ex\left\vert #1 
    \right\vert\kern-0.25ex\right\vert\kern-0.25ex\right\vert}}
\newcommand{\tripl}[1]{{\vert\kern-0.25ex\vert\kern-0.25ex\vert #1 
    \vert\kern-0.25ex\vert\kern-0.25ex\vert}}

\begin{document}


\title{Determining an asymptotically AdS Einstein spacetime from data
  on its conformal boundary}

\author{Alberto Enciso}
 \email{aenciso@icmat.es.}
 \affiliation{ Instituto de Ciencias Matem\'aticas, Consejo Superior de
  Investigaciones Cient\'\i ficas, 28049 Madrid, Spain
}%

\author{Niky Kamran}
 \email{nkamran@math.mcgill.ca.}
\affiliation{Department of Mathematics
  and Statistics, McGill University, Montr\'eal, Qu\'ebec, Canada H3A 2K6}%
\date{\today}

\begin{abstract}
  An outstanding question lying at the core of the AdS/CFT
  correspondence in string theory is the holographic prescription
  problem for Einstein metrics, which asserts that one can slightly
  perturb the conformal geometry at infinity of the anti-de Sitter
  space and still obtain an asymptotically anti-de Sitter spacetime
  that satisfies the Einstein equations with a negative cosmological
  constant. The purpose of this paper is to address this question by
  providing a precise quantitative statement of the real-time
  holographic principle for Einstein spacetimes, to outline its proof
  and to discuss its physical implications.
\end{abstract}

\maketitle

Our purpose in this Letter is to present the precise quantitative statement of a well-posedness result for the Einstein equations with negative cosmological constant, in $n$ dimensions, with boundary data corresponding to a small perturbation of the $n$-dimensional anti-de-Sitter (AdS) geometry at time-like conformal infinity. We will set the physical and mathematical context for our result and sketch the main ideas of the proof, referring the reader to~\cite{IM} for the analytical details and precise estimates needed to prove our claim. 
Our theorem provides a mathematical rigorous illustration of the so-called holographic principle, which asserts that a Lorentzian
conformal metric $\hg_{\mu\nu}$ on the boundary propagates through a
suitable $(n+1)$-dimensional bulk spacetime and gives rise to an
Einstein metric $g_{\mu\nu}$ (typically a small perturbation of the
AdS spacetime) by solving the Einstein equations with a negative
cosmological constant. 
Independently of its mathematical interest as a general well-posedness theorem for the Einstein equations with data prescribed at time-like conformal infinity, part of the motivation for this work originates in the AdS/CFT correspondence, an important relation
in string theory first proposed by Maldacena~\cite{Maldacena}, which
establishes a connection between conformal field theories in $n$
dimensions and gravity fields on an $(n+1)$-dimensional spacetime of
anti-de Sitter type, to the
effect that correlation functions in conformal field theory are given
by the asymptotic behavior at infinity of the supergravity
action. This implies that the gravitational field in
$(n+1)$-dimensions must be holographically determined by the
conformal field, which plays the role of boundary data on the
(timelike) conformal
boundary on an asymptotically AdS spacetime that is referred to as the
bulk. It is remarkable, though, that nearly all the key references on the AdS/CFT
correspondence, ever since Witten~\cite{Witten}, have implemented the
concept of holographic prescription using the Euclidean (or
Wick-rotated) version of the AdS spacetime, that is, the hyperbolic
space. Let us emphasize at this point, however, that there are a few
important papers available on the implementation of the AdS/CFT
correspondence in Lorentzian
signature\cite{many,many2}, which,
focusing on linearized gravity theories, reconstruct the bulk metric
from boundary data such as the expectation of the stress-energy tensor
of the boundary theory, where the ambiguities corresponding to the
conformal anomalies have been computed. Our result provides a stepping stone 
to the implementation of this program in the fully non-linear setting where the space-time geometry 
is obtained by solving the Einstein equations with asymptotically AdS boundary data. 

Our goal in this paper is thus to develop a full-fledged holographic
principle for Einstein gravity, thereby addressing a well-known open question in
the literature. The importance of this is twofold. Firstly, on a fundamental
level, it is well-known that the physical
nature of the Einstein equations in Euclidean signature is inherently different from that of
the real-time problem. Indeed, in Lorentzian signature the Einstein
equations describe the metric in terms of nonlinear waves that
determine the geometry of the spacetime as they propagate, while the
Euclidean signature describes an equilibrium configuration where there
is no time evolution present. 

Secondly, from a practical standpoint, it is essential to understand
the folk wisdom that the Wick-rotated problem is perfectly good for most formal
computations. We will see later on that the rationale behind this
statement is quite simple: the spacetime metric can be written as the
sum of two factors, one of which is ``large'' at the conformal boundary and the
other is ``small''. The key feature is that the large term is
obtained from the boundary data using algebraic methods, which behave
well with respect to Wick rotations. What is highly nontrivial,
though, is that the small term at infinity (which is by no means small
elsewhere in the bulk) does not depend on the boundary data in an
algebraic fashion, but rather it must be carefully crafted using
functional-analytic techniques that do not behave well at all under
Wick rotations. In particular, this part of the spacetime metric is
not related in any direct way with its Euclidean-signature
counterpart. The redeeming feature of this decomposition is that, when
one is interested in the asymptotic behavior of the metric at infinity
and its connections with the correlation functions of the associated
conformal field theory on the boundary, the leading terms in the
asymptotic expansion depend only on the part of the metric that is
large at infinity, and are therefore granted to agree with the Wick-rotated
contributions of the Euclidean version of the problem.

Let us elaborate on the formulation of the holographic principle. For the
Euclidean version of the problem, it has been known since
Witten~\cite{Witten} that the holographic principle follows from a
theorem in Riemannian geometry, due to Graham and Lee~\cite{GL}, that
asserts that one can ``tickle'' the conformal geometry at infinity of
the hyperbolic space and still obtain an Einstein metric on the
ball. More precisely, the Graham--Lee
theorem says that if $\hg$ is a metric on the $n$-dimensional sphere $\mathbb{S}^{n}$ that is
close enough to the canonical sphere metric in some suitable norm,
then there is a Riemannian metric~$g_{\mu\nu}$ in the
$(n+1)$-dimensional ball $\BB^{n+1}$ that satisfies
the Einstein equation
\begin{equation}\label{Einstein}
R_{\mu \nu}=-\La g_{\mu\nu}
\end{equation}
with the property that the rescaled metric $\Bg:=x^2 g$ is continuous
up to the boundary and satisfies the boundary condition
\begin{equation*}
\Bg|_{T\pd\BB^{n+1}}=\hg\,,
\end{equation*}
possibly up to multiplication by a positive scalar function reflecting
the conformal invariance of the boundary condition. Here $x$ is any
positive function on the ball which vanishes to first order on the
boundary and we are denoting by
$\Bg|_{T\pd\BB^{n+1}}$ the pullback of the metric $\Bg$ to the
boundary. The Graham--Lee theorem also guarantees that the metric $g_{\mu\nu}$
is close to that of the hyperbolic space is some suitable sense.

As is well-known, once one has used the Graham--Lee theorem to ensure that the Einstein
metric~$g$ corresponding to the boundary data is actually well
defined, the results of Fefferman and
Graham~\cite{FG} provide precise quantitative information on the
asymptotic behaviour of the Euclidean Einstein metric near the
boundary. As shown by Witten~\cite{Witten}, this makes it
possible to compute the divergent terms that arise from regularizing
the gravitational action and relating it to the partition function of
the conformal field theory on the boundary, giving rise to a local
expression for quantities such as the conformal anomaly. 

The reason for which the Graham--Lee theorem cannot be readily
extended to the real-time (or Lorentzian) situation is directly related with the
change of the underlying physics that the Einstein equation
describes. In the Euclidean case, the Einstein equations are
essentially a system of elliptic partial differential equations, of
the kind describing equilibrium configurations of a physical
system. Just as the Laplacian corresponds to
the wave operator under Wick rotation, in the real-time setting the
Einstein equations are a system of nonlinear wave equations, which are
much harder to handle and, in an asymptotically AdS spacetime, present
the unpleasant feature of being strongly singular at
infinity. Therefore, proving the holographic principle in Lorentzian
signature involves solving a hard mathematical problem with a large
body of related literature that covers waves in AdS\cite{AdS}
backgrounds and
linear~\cite{linear} and nonlinear~\cite{nonlinear} waves in
asymptotically AdS backgrounds. 

We would like to mention at this stage that Friedrich~\cite{Friedrich} has proved an existence theorem for asymptotically simple solutions of the four-dimensional Einstein equations with negative cosmological constant using a different approach based on a conformal representation for the Einstein
equations obtained through the use of normal conformal connections. The analysis is carried out in~\cite{Friedrich} under stronger regularity assumptions, in which the initial and boundary data are assumed to be smooth, giving rise to solutions admitting a smooth conformal boundary in a neighbourhood of the initial three-dimensional Cauchy hypersurface. Our approach based on the Graham-Lee formulation of the Einstein equations, besides being valid in $n$ dimensions, deals with data of lower regularity, and gives a quantitative statement of well-posedness for data close to the AdS boundary data at infinity.

We shall next get into the details of the real-time holographic
principle for the Einstein equations. To this end, we will outline the proof of the fact that
any suitably small perturbation of the conformal geometry at infinity
of the anti-de Sitter space $\AdS$ gives rise to an Einstein
metric and discuss the salient features of the
construction. Inessential technicalities will be avoided throughout.

Let us begin by recalling the notation for the $(n+1)$-dimensional AdS
spacetime, which is given by the metric
\[
\gads= \frac{-(1+x^2)^2\,d t^2+d x^2+(1-x^2)^2\, d\Om_{n-1}^2}{x^2}
\]
in the the solid cylinder $\RR\times \ball$. We have rescaled the
metric in order to choose the cosmological constant
$\La:=n$, which makes the expressions slightly more manageable. In the
expression for the metric,
$d\Om_{n-1}^2$ is the standard metric on the $(n-1)$-dimensional
sphere and the variable $x$ takes values in $(0,1]$.
It is thus manifest that the rescaled metric $\bgads:=x^2\gads$ is
well behaved up  to the
boundary $\RR\times\pd\ball$, which describes the timelike infinity of
the spacetime.  The conformal infinity corresponds to $x=0$, so it is
given by the conformal metric
\[
\gcyl:=-dt^2+d\Om_{n-1}^2
\]
is the cylinder $\mathbb{R}\times
\mathbb{S}^{n-1}$. 

As before, what we need to prove is that one can tickle the AdS
spacetime at infinity and still obtain an Einstein
metric. Specifically, we aim to show that if we take a Lorentzian
metric~$\hg$ on the cylinder that is a small perturbation of the above
metric~$\gcyl$ in some suitable sense, then there is a spacetime
metric~$g$ that satisfies the Einstein equations~\eqref{Einstein} and
whose conformal infinity is given by~$\hg$, namely,
\begin{equation}\label{BC}
\Bg|_{T(\RR\times\pd\ball)}=\hg\,,\qquad \text{with } \Bg:=x^2 g\,.
\end{equation}
Furthermore, since the Einstein
equations in Lorentzian signature are wave equations, we also specify initial conditions for
the metric, which we can write as
\begin{equation}\label{inic}
g|_{t=0}=g_0\,,\qquad \pd_t g|_{t=0}=g_1\,,
\end{equation}
It is well-known that the admissible choices of $g_0$ and $g_1$ are
subject to the constraints equations of general relativity. Additionally, the initial and boundary conditions
$(g_0,g_1,\hg)$ must satisfy certain nontrivial compatibility
conditions. 

For the purposes of this paper, we will state the result in the
simplest case, namely, trivial initial data and
nontrivial boundary data: $g_0:=\gads$, $g_1:=0$ and $\hg$ any
Lorentzian metric that is identically equal to $\gcyl$ for
$t\leq0$. This corresponds to the physically relevant situation of how
the spacetime departs from the AdS spacetime metric as nontrivial boundary
data are switched on at the conformal infinity at time $t=0$. In the general case,
the result is qualitatively the same, but the precise smallness assumptions on the initial
conditions are cumbersome and not particularly illuminating. Full 
details can be found in~\cite{IM}.

Specifically, the result that establishes the holographic principle for the
Einstein equations can be quantitatively stated as follows:
For any spatial dimension $n\geq 3$, let $\hg$ be a Lorentzian metric
on the cylinder $\mathbb{R}\times \es$ that is a small perturbation of
the canonical one, $\gcyl$. More precisely, we assume that the difference
$\hg-\gcyl$ and all its spacetime derivatives up to order $3n+9$ are bounded by a
small constant, which we write as
\[
\|\hg-\gcyl\|_{C^{3n+9}(\RR\times\es)}<\de\,.
\]
If $\de$ is small enough, then for some time $T$ of order $1/\de$
there is an asymptotically AdS metric $g$ that satisfies the Einstein
equation~\eqref{Einstein} on
$(-T,T)\times\ball$ and the above boundary and initial
conditions~\eqref{BC}-\eqref{inic}. Furthermore, the Einstein
metric~$g$ and the exact AdS metric are close in the sense that
\[
\|\bar g-\bgads\|_{C^{n-2}((-T,T)\times\ball)}<C\de
\]
for some constant $C$ that only depends on~$n$.

A couple of comments are now in order. First and foremost it should be
noticed that, in stark contrast with the Euclidean version of the
problem, we can only grant the existence of the Einstein metric for
some finite time $T$ because, in keeping with the generally held view
that anti-de Sitter space should be non-linearly unstable, black holes
are expected to appear. However, this does not invalidate previous results,
obtained using linearized real-time formulations~\cite{many}, because
the time that these singularities take to form is at least 
$c/ \de$, with $c$ an absolute constant, and thus becomes very large
for small perturbations of the conformal infinity. Secondly, as is well-known, the fact that up
to $n-2$ derivatives of the difference $\Bg-\bgads$ are small is
optimal in view of the Fefferman--Graham expansion for the
metric $g$, in which logarithmic terms appear starting with
$x^{n-1}\log x$. On the contrary, we have not strived to provide a
sharp bound for the number of derivatives of $\hg$ that must be
close to $\gcyl$.

We shall next outline the main ideas of the proof of this result,
skipping all the technicalities for the benefit of the reader. The first step in the proof is to remove
the gauge freedom that arises from the diffeomorphism invariance of
the Einstein equation. Since the classical work of Choquet-Bruhat, a standard way of doing
this has been to use wave coordinates, but in this case it is more convenient to use a
generalization of this idea that is known as DeTurck's
trick~\cite{DeTurck2}. It allows us to modify the Einstein
equation to an equivalent, less degenerate form that is indeed given
by a system of nonlinear wave equations.
 
To apply DeTurck's trick, we choose a reference metric $\gref$ to be
an asymptotically AdS metric whose pullback to the conformal boundary
is $\hat g$. We refer to~\cite{IM} for the construction of this
metric, which is essentially obtained by interpolating (in~$x$) the
AdS metric and the asymptotic metric $(dx^2+\hg)/x^2$, which one can easily make
sense of in a neighborhood of the boundary.  Let us denote by
$\Ga^\nu_{\la\rho}$ and $\widetilde\Ga^\nu_{\la\rho}$ the Christoffel
symbols of the metrics~$g$ and~$\gref$, respectively. DeTurck's trick
consists in looking for solutions to the modified Einstein equation
\begin{equation}\label{eqQ}
Q_{\mu\nu}=0\,,
\end{equation}
where the tensor $Q\equiv Q(g)$ is
\begin{equation}\label{Q}
Q_{\mu\nu}:= R_{\mu\nu}+n g_{\mu\nu}+\frac12(\nabla_\mu W_\nu+\nabla_\nu W_\mu)\,.
\end{equation}
Here the covariant derivatives and the Ricci tensor are computed with respect of the
metric $g$ and 
\begin{equation}\label{W}
W_\mu:=g_{\mu\nu}\,g^{\la\rho}\,(\Ga^\nu_{\la\rho}-\widetilde\Ga^\nu_{\la\rho})\,.
\end{equation}
Notice that~$Q_{\mu\nu}$ depends on the initial and boundary conditions through the
reference metric~$\gref$. The point now is that, choosing suitable
initial conditions, solving the equation~\eqref{eqQ}
is tantamount to solving the Einstein equation~\eqref{Einstein}, but
the former has much better solvability properties as it is not gauge invariant.

Our goal now is to solve the modified Einstein equation~\eqref{eqQ}
together with the above initial and
boundary conditions~\eqref{BC}-\eqref{inic}.
For metrics that are asymptotically AdS at infinity, of the kind that we have in
this problem, the coefficients of this equation are strongly singular at
$x=0$. Symbolically, Eq.~\eqref{eqQ} reads as
\begin{equation}\label{Qsing}
\Bg^{\al\be}\pd_\al\pd_\be g_{\mu\nu}+
\frac{1}{x}A^{\al\be\la}_{\mu\nu}\pd_\la \Bg_{\al\be} + \frac{1}{x^2} B^{\al\be}_{\mu\nu}\Bg_{\al\be} =0\,,
\end{equation}
where the tensors $A$ and $B$ depend on $\Bg,\pd\Bg$.


The second step is to construct an approximate solution~$\ga$ of the
modified Einstein equation~\eqref{eqQ} with the desired behavior at
the conformal boundary. This will be the part of the solution that is
large at infinity, in the sense that we informally discussed above. This approximate solution can be computed
algorithmically and has the key feature that it can be obtained using only
local information about~$\hg$ (in fact, using only derivatives and
algebraic operations), so it is well behaved under
Wick rotation. An important consequence is that, in particular, the
approximate solution~$\ga$ would not have ``seen'' the initial
conditions $g_0$ and $g_1$, if had taken nontrivial ones, and this is
also important because actually in the Wick-rotated situation one cannot impose
any initial conditions. This explains why, to a high asymptotic order
at the boundary, the spacetime metric is independent of the choice of
initial conditions, which justifies that one often forgets about them in
the description of the holographic principle and in its applications
to computing quantities for concrete physical theories.

The approximate solution~$\ga$ is constructed by
successively ``peeling off'' the leading order behavior of the metric~$g$
at the conformal boundary, in a manner that is formally similar to the Riemannian case. This recursive procedure is described in
detail in~\cite{IM}, and can be summarized as follows. 
Given nonnegative integers $s$ and $\si$, we say that a
symmetric tensor field $q$ is in $\cO_j(x^s\log^{\leq \si}x)$ if it can be
written in the form 
\[
q=x^s\sum_{\si'=0}^\si(\log x)^{\si'} B^{\si'}\,,
\]
where $B^{\si'}$ is a smooth symmetric tensor field in $I\times\oball$
satisfying the bounds
\[
\|B^{\si'}\|_{C^k(I\times\ball)}\leq
F_{k}\big(\|\hg\|_{C^{k+j}(I\times\es)}\big)
\]
for each $k$ and $r$, provided that the difference
\[
\|\hg-\gcyl\|_{C^{0}(I\times\es)}
\]
is small
enough. Here $F_{k}$ is a polynomial with $F_{k}(0)=0$. 
It is important to note that in all the terms of
the form $\cO_j(x^s\log^{\leq \si} x)$, the coefficients of the corresponding polynomials $F_{k}$ are bounded
independently of~$\hg$. The peeling behaviour can then be described as follows:
Let $l$ be a nonnegative integer and suppose that
\[
\|\hg-\gcyl\|_{C^p(I\times\es)}<\de
\]
for some small enough $\de>0$ and $p\geq l$. Then there is a weakly
asymptotically AdS metric $\ga_l$ on $I\times\ball$ of the form
\begin{equation}\label{gal}
\ga_{l}=\sum_{k=0}^l \cO_j(x^{k-2}\log^{\leq \si_k}x)
\end{equation}
whose pullback to the boundary of $\Bga_l:=x^2\ga_l$ is
\[
(j_{I\times\pd\ball})^* \Bga_l=\hg
\]
and such that 
\begin{equation}\label{Qgal}
Q(\ga_l)=\cO_{l+1}(x^{l-1}\log^{\leq \si_l}x)+ \cO_{l+2}(x^l\log^{\leq \si_l}x)\,,
\end{equation}
where $\si_k$ is a nonnegative integer that is equal to zero for all
$k\leq n-2$. Furthermore, the metric $\ga_l$ is close to $\gads$ in the sense that
\begin{equation}\label{boundgl}
\|\Bga_l-\bgads\|_{C^{p'}_r(I\times\ball)}<C\de
\end{equation}
with $p':= \min\{p-l, n-2\}$ and $r:=p-l-p'$, while $Q(\ga_l)$ is bounded by
\begin{equation}\label{boundQgal}
\|x^{2-l}Q(\ga_l)\|_{C^0_{p-l-2}(I\times \ball)} <C\de\,.
\end{equation}
Here $\|\cdot\|_{C^{p'}_r(I\times\ball)}$ denotes the supremum  norm of the
space of functions on $I\times\ball$ with $m+r$ continuous derivatives,
with the peculiarity that each of the last $r$ derivatives with respect to~$x$ is
regularized by multiplying by~$x$. This way, for instance, for all
$k,l,m\geq 1$ we have that
\begin{equation}\label{xlogxklm}
x^{m}\, (\log x)^l
\end{equation}
is in $C^{m-1}_k$ but not in $C^{m+k-1}$. The rigorous definition of
this norm can be found in~\cite{IM}.


The final and key step in the argument is to show that the approximate
solution~$\ga$, which we have obtained using essentially algebraic
means, can be promoted to an actual solution~$g$ of the modified Einstein
equation~\eqref{eqQ}. This is a functional-analytic argument that is based on the 
convergence of an iterative procedure and, as
such, is not robust 
under Wick rotation. As we shall see below, the idea is to use an iterative argument
where we decompose the metric as
\begin{equation}\label{gu}
g=\ga + x^{\frac n2}u\,,
\end{equation}
where
\[
\ga\equiv \ga_l
\]
is the metric  $\ga_l$ constructed constructed above, with some large
enough value of the parameter~$l$.
This makes precise the idea that the asymptotically AdS metric~$\ga$ is
the part of the metric that is large at the boundary and the other
term, which is convenient to write as $x^{\frac n2}u$, is
smaller; precise estimates are given in~\cite{IM}.

We now describe in greater detail the steps leading to a solution of the equation $Q(g)=0$ with the
desired initial and boundary conditions. To this end, let us write the
solution as
\[
g=:\ga+ h\,.
\]
Let us recall from Eq.~\eqref{Qsing} that one can write $Q(g)$ in local
coordinates as
\[
Q(g)=\widetilde P_gg+B(g)\,,
\]
where~$\widetilde P_g$ is the linear differential defined by
operator  as
\[
(\widetilde P_gg')_{\mu\nu}:=-\frac12 g^{\la\rho}\pd_\la\pd_\rho g_{\mu\nu}'
\]
and $B(g)$ depends on $g$ and quadratically on $\pd g$. Taylor's
formula ensures that
\begin{equation}\label{Qg1}
B(g)=B(\ga)+(DB)_\ga h-\tcE(h)\,,
\end{equation}
where the error term is
\begin{equation}\label{tcE}
\tcE(h):=-\int_0^1(D^2B)_{\ga+sh}(h)\, ds
\end{equation}
and the second order differential of $B$ is understood as a quadratic
form. The equation $Q(g)=0$ can then be written as
\begin{equation}\label{Qg2}
\widetilde P_gh+(DB)_gh+(\widetilde P_\ga\ga-\widetilde P_g\ga)+Q(\ga)-\tcE(h)=0\,.
\end{equation}

We shall next write the equation $Q(u)=0$ in a way that is easier to
solve using an iterative scheme. For this, let us begin by defining a linear operator $T_{g}$ as
\[
T_gh:=-3h(\nabla^{(\ga)}x, \nabla^{(\ga)}x)\, \Bg\,,
\]
where $\nabla^{(\ga)}$ stands for the connection associated with the
metric~$\ga$. As easy computation shows that $T_g$ is the
differential of the function 
\[
g\mapsto \widetilde P_\ga\ga-\widetilde P_g\ga
\]
at the point $g=\ga$. Hence we will set
\begin{equation}\label{tcF}
\tcF(h):=T_gh+\widetilde P_g\ga-\widetilde P_\ga\ga\,,
\end{equation}
which plays then the role of a quadratic correction term. In view of~\eqref{Qg2}, the equation
$Q(g)=0$ then reads as
\[
\widetilde P_gh+(DB)_gh+T_gh=-Q(\ga)+\tcF(h)+\tcE(h)\,.
\]
To conclude the reformulation of the equation, let us introduce a new
unknown $u$ as
\[
h=:x^{\frac n2}u\,.
\]
In terms of~$u$, it is natural to consider the $g$-dependent linear differential operator $P_g$
given by
\[
\widetilde P_gh+(DB)_gh+T_gh=:x^{\frac n2+2}P_gu\,,
\]
which finally enables us to write the equation
$Q(g)=0$ as
\begin{equation}\label{EinsteinP}
P_gu=\cF_0+\cG(u)\,.
\end{equation}
Here $\cF_0$ and $\cG$ are given by
\[
\cF_0:=-x^{-\frac n2-2}Q(\ga)\,,\qquad \cG(u):=\cF(u)+\cE(u)\,,
\]
with
\[
\cF(u):=x^{-\frac
  n2-2}\tcF(h)\,,\qquad \cE(u):=x^{-\frac n2-2}\tcE(h)\,.
\]

It is shown in~\cite{IM} that his equation may be solved using an iterative
procedure that will produce $u$ as the limit of a sequence $u^m$,
with $u^1:=0$ and 
\begin{equation*}
P_{g^{m}}u^{m+1}=\cF_0+\cG(u^{m})\,.
\end{equation*}
Of course, here $g^m:= \ga+ x^{\frac n2}u^m$ and the initial conditions
that we need to impose are
\begin{equation*}
u^{m+1}|_{t=0}=u_0\,,\qquad \pd_t u^{m+1}|_{t=0}=u_1\,,
\end{equation*}
where we have set 
\begin{equation}\label{u01}
u_j:=x^{-\frac n2}(g_j-\pd_t^j\ga|_{t=0})\qquad \text{for }j=0,1.
\end{equation}

For this, we set up the iteration that will produce the function $u$
as the limit of a sequence $u^m$ as $m\to\infty$. The metric
associated to $u^m$ via~\eqref{gu} will be denoted
by~$g^m$. 
Written in the form~\eqref{EinsteinP}, the iteration consists in starting
from $u^1:=0$ and obtaining $u^{m+1}$ recursively by solving the
linear problem
\begin{equation*}
P_{g^{m}}u^{m+1}=\cF_0+\cG(u^{m})
\end{equation*}
with zero initial and boundary conditions. Details are given in~\cite{IM}, where it is also proved
that the contribution of the term $x^{\frac n2} u$ to the asymptotics
of the metric~$g$ at the conformal boundary are of order $x^n$
(provided that the parameter~$l$ appearing in the peeling expansion~(\ref{gal}) is
chosen large enough). This
shows why the ``algebraic'' part~$\ga$ is the only one that we need to
compute the first terms in the expansion of~$g$ at $x=0$.

We conclude this paper with a few remarks and perspectives. The main
result of our paper provides an answer to the holographic prescription
problem in the form of an existence theorem valid in in any space dimension $n\geq 3$ for Einstein metrics with prescribed conformal infinity consisting in suitably small perturbation of the conformal infinity of anti-de Sitter space. From the knowledge of the leading order behaviour of the Einstein metric near the conformal boundary, it is possible in principle to estimate quantities of physical interest for the admissible boundary conformal field theories, such as two-point or $n$-point correlation functions~\cite{Maldacena, Witten}. The computation of conformal anomalies~\cite{Witten, many2} would be of great interest in the Lorentzian setting.

From the point of view of the Cauchy problem for the Einstein
equations, the existence of the metric~$g$ solving the holographic
prescription problem is only local in time, in keeping with the fact
that anti-de Sitter space is expected to be non-linearly
unstable. Our estimates show that the interval of definition of the time parameter is of order $1/\delta$, where $\delta$ is the size of the deviation of the boundary data from the AdS boundary data. It is a very challenging goal to devise a
rigorous scenario for singularity formation.

A.E.\ is supported by the ERC Starting Grant 633152 and thanks McGill University for hospitality
and support. A.E.'s research is supported in part by the ICMAT Severo
Ochoa grant SEV-2011-0087 and the MINECO grant  FIS2011-22566. The research of N.K.\ is supported by NSERC grant
RGPIN 105490-2011.
\bibliography{aipsamp}

\begin{thebibliography}{99}\frenchspacing
%


\bibitem{Maldacena}
J. Maldacena, Adv. Theor. Math. Phys. 2 (1998) 231.


\bibitem{Witten}
E. Witten, 
Adv. Theor. Math. Phys. 2 (1998) 253. 

\bibitem{many}
V. Balasubramanian, P. Kraus, Comm. Math. Phys. 208 (1999) 413;
D.T. Son, A.O. Starinets, JHEP 09 (2002) 042.

\bibitem{many2}
S. de Haro, S.N. Solodukhin, K. Skenderis, Comm. Math.Phys. 217 (2001)
595; M. Bianchi, D.Z. Freedman, K. Skenderis,  Nucl.Phys. B631 (2002)
159.

\bibitem{GL}
C.R. Graham, J.M. Lee, Adv. Math. 87 (1991) 186.


\bibitem{FG}
C. Fefferman, C.R. Graham, {\em The ambient metric}, Princeton University Press, Princeton, 2012.


\bibitem{AdS}
P. Breitenlohner, D.Z. Freedman, Ann. Physics 144 (1982) 249;
Y. Choquet-Bruhat, Class. Quant. Grav. 6 (1989) 1781; A. Ishibashi, R.M. Wald, Class. Quant. Grav. 21 (2004) 2981;
 A. Bachelot, J. Math. Pures Appl. 96 (2011) 527; A. Enciso, N. Kamran, {Phys. Rev.~D} 85 (2012) 106016;
 A. Bachelot, 
Comm. Math. Phys. 320 (2013) 723.

\bibitem{linear}
A. Vasy, Anal. PDE 5 (2012) 81;
G. Holzegel, J. Hyperbolic Differ. Equ. 9
(2012) 239; C. Warnick,
Comm. Math. Phys. 321 (2013) 85; G. Holzegel, C. Warnick,
J. Funct. Anal. 266 (2014) 2436.

\bibitem{nonlinear}
A. Enciso, N. Kamran, J. Math. Pures Appl. 103 (2015) 1053.




\bibitem{Friedrich}
H. Friedrich, J. Geom. Phys. 17
(1995) 125--184.





\bibitem{DeTurck2}
D.M. DeTurck, Compositio Math. 48 (1983) 327.



\bibitem{IM} A. Enciso, N. Kamran, Lorentzian  Einstein metrics with
  prescribed conformal infinity, arXiv:1412.4376.





\end{thebibliography}

\end{document}